\documentclass[a4paper,10pt]{article}
	\usepackage{enumerate}
	\usepackage{color}
	\usepackage[utf8]{inputenc} 
	\usepackage[english]{babel}
	\usepackage[T1]{fontenc}
	\usepackage{graphicx}
	\usepackage{amsfonts,amssymb,amsmath,latexsym,amsthm}
	\usepackage{textcomp}
        \usepackage{tikz}
	\usepackage[pdftex]{hyperref}
	\usepackage{geometry}
	\usepackage{tikz}
	\DeclareMathOperator{\sech}{sech}
	\DeclareMathOperator{\sign}{sign}

	\author{M. V. Flamarion$^1$, T. Gao$^2$ \&  R. Ribeiro-Jr$^3$}
	\title{An investigation of the flow structure beneath solitary waves with constant vorticity on a conducting fluid under normal electric fields}

	\date{}
	
\begin{document}
	\maketitle
	\begin{center}
		
		{\footnotesize $^{1}$ Unidade Acad{\^ e}mica do Cabo de Santo Agostinho, 
	UFRPE/Rural Federal University of Pernambuco, BR 101 Sul, Cabo de Santo Agostinho-PE, Brazil,  54503-900 \\
	marcelo.flamarion@ufrpe.br }

	\vspace{0.3cm}
	
	{\footnotesize $^{2}$ Department of Mathematical Sciences, University of Essex, Colchester CO4 3SQ, UK.\\ t.gao@essex.ac.uk}

		\vspace{0.3cm}

		{\footnotesize $^{3}$ UFPR/Federal University of Paran\'a,  Departamento de Matem\'atica, Centro Polit\'ecnico, Jardim das Am\'ericas, Caixa Postal 19081, Curitiba, PR, 81531-980, Brazil \\ robertoribeiro@ufpr.br}
		
%	\vspace{0.3cm}
%		{\footnotesize $^{4}$ Department of Mathematical Sciences, University of Bath, Bath BA2 7AY, UK.\\ add49@bath.ac.uk }	
	
	%{\footnotesize ORCID Number: 0000-0002-0877-4831}

	\end{center}

	%\maketitle
	
	\begin{abstract} 
	\noindent  The motion of an interface separating two fluids under the effect of electric fields is a subject that has picked the attention of researchers from different areas. While there is an abundance of studies investigating the free surface wave properties, very few works have examined the associated velocity field within the bulk of the fluid. Therefore,  in this paper, we investigate numerically the flow structure beneath solitary waves with constant vorticity on an inviscid conducting fluid bounded above by a dielectric gas under normal electric fields in the framework of a weakly nonlinear theory.  	Elevation and depression solitary waves with constant vorticity are computed by a pseudo-spectral  method and a parameter sweep on the intensity of the electric field is carried out in order to study  its role in the appearance of stagnation points. We find that for elevation solitary waves  the location of stagnation points does not change significantly with  variations of the electric field. For  depression solitary waves, on the other hand, the electric field acts as a catalyser that makes possible the appearance of stagnation points -- in the sense that in its absence there is no stagnation point.

%	Waves with constant vorticity and electrohydrodynamics flows are two topics in fluid dynamics that have attracted much attention from scientists for both the mathematical challenge and their industrial applications.  The coupling of electric fields and vorticity is of significant research interest. In this paper, we study the flow structure of steady periodic travelling waves with constant vorticity on a dielectric fluid under the effect of normal electric fields.  Through the conformal mapping technique combined with pseudo-spectral numerical methods, we develop an approach that allows us to conclude that the flow can have zero, two or three stagnation points according to variations in the voltage potential. We describe in detail the recirculation zones that emerge together with the stagnation points. Besides, we show that the number of local maxima of the pressure on the bottom boundary is intrinsically connected to the saddle points. 

		\end{abstract}

	\section{Introduction}
 Electrohydrodynamics (EHD) is an interdisciplinary subject that studies the coupling of  fluid dynamics and electromagnetism. The motivation comes from the engineering applications of manipulating fluid motion by electric fields. The readers may refer to \cite{CCY} for more details. 

An EHD problem is usually concerned with an interface  between two fluids, and therefore the fluid motion under the effect of electric fields is governed by the Navier-Stokes equations (or the Euler equations in the inviscid case) coupled with Maxwell's equations. A complete review has been produced by Papageorgiou \cite{D}.

The motion of a free surface wave in an EHD  flow has been widely studied by different frameworks. Of note, many reduced models have been derived for different configurations under certain assumptions, such as long-wave approximations in order to understand the mechanism of fluid-electric  coupling. The readers may refer to \cite{DGVK,Wang} for a comprehensive review of the linear theory and  the weakly nonlinear theory respectively.   However, very few works have focused  on the features of a velocity  field associated with free surface  waves in EHD flows. To our knowledge, the only study in this direction is the one carried out by Flamarion {\em et al.} \cite{FGRD:2022} in which the authors show that normal electric field acts as a mechanism that helps  the appearance of stagnation points beneath periodic waves with constant vorticity.  Stagnation points can be understood as points in the fluid domain that travels at the same speed as the wave.

No work has been achieved to study the flow structure beneath solitary waves under electric fields to our best knowledge. To fill such a gap, we consider the same configuration as in Gleeson {\em et al.} \cite{GHPV} who derived  a Korteweg-de Vries Benjamin-Ono equation to describe the fluid interface.  Then we compute numerically solitary waves with constant vorticity and investigate the electrical effect on the streamlines.  We shall focus on comprehending the role of the electrical field in the appearance of stagnation points.   

 We will proceed  using  Korteweg-de Vries Benjamin-Ono equation to approximate the velocity  field in the bulk of the fluid   and then extract  information about the flow structure  beneath the wave. This methodology of approximating the velocity field through reduced models has been adopted in other studies such as in irrotational gravity flows  \cite{BK,G,Khorsand}, in gravity flows with constant vorticity  \cite{Guan:2020,AK,CK,CK2},  in capillary-gravity flows with constant vorticity   \cite{Wave Motion:2023, TEMA:2023} and in gravity flows with variable vorticity  \cite{FR2023}. There is no doubt that solving the full Euler equations provides a more complete description of the flow. However, reduced models can reproduce the main features  of the flow with little computational effort. 

The  paper is structured as follows. We recall the formulation in section \ref{fm}. The numerical methods are introduced in section \ref{ns} and \ref{pt}. The results are presented in section \ref{res}.
	\section{Mathematical Formulation}\label{fm}

\begin{figure}[!h]
    \centering
\begin{tikzpicture}
%tikz picture content goes here
\draw [ ->,thick] (8,2) -- (8,4) ;
	\draw [ ->,thick] (4,2) -- (4,4) ;
	\draw [ ->,thick] (5,2) -- (5,4) ;
 	\draw [ ->,thick] (6,2) -- (6,4) ;
	\draw [ ->,thick] (7,2) -- (7,4) ;
        \draw [ -,thick] (0,-1) -- (10,-1);
        		\draw [fill=gray,gray] (0,-1) rectangle (10,-1.05); 
        \node[text width=2cm] at (-.1,-1) {\small{$y=-1$}};
	\node[text width=4cm] at (7.4,3.2) {\small{$E$}};
	
	\draw [ -,line width=1pt] (0,1) -- (10,1)[dashed] ;
	\draw[thin]  plot [samples=300,domain=0.1:10] (\x,{1+.3*sin(2*(\x +pi/2-pi-.95) r)/2+.3*sin((\x -pi+pi/2-.95) r)/2});
	\node[text width=2cm] at (-.1,1) {\small{$y=0$}};
	%\node[text width=2cm] at (7,2) {\small{$\nabla^2v=0$}};
	\node[text width=4cm] at (7,-.5) {\small{$v=0$}};
	\node[text width=4cm] at (3,-.2) {\small{Conductor}};
	\node[text width=4cm] at (3,1.8) {\small{Permittivity} $\epsilon_d$};
	\node[text width=4cm] at (3,2.2) {\small{Dielectric}};
	\node [text width=2cm] at (8,0.5) {\small{$y=\zeta (x,t)$}};

	\draw [ ->,thick]  (9.5,.5) -- (9.5,-.5)  node [right] {\small{$g$}};
	\draw [ ->,thick]  (.5,3) -- (.5,3.5)  node [left] {\small{$y$}};
	\draw [ ->,thick]  (.5,3) -- (1,3)  node [right] {\small{$x$}};
	
\end{tikzpicture}%
   
    \caption{Schematic of the problem.}
    \label{fig:scm}
\end{figure}
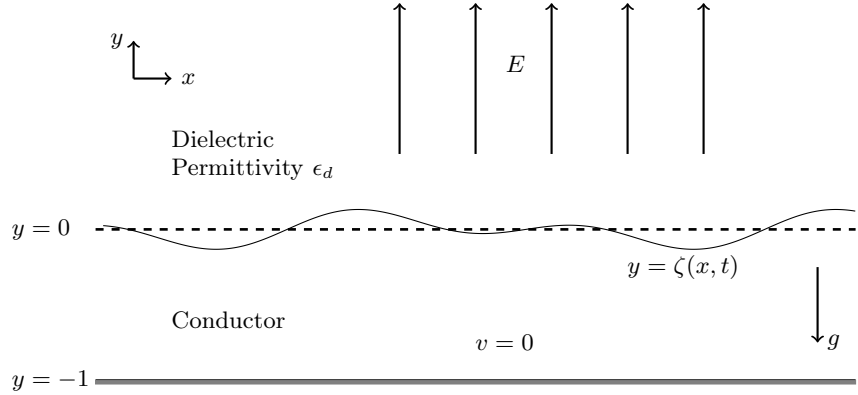

	We consider an incompressible flow of an inviscid conducting fluid of constant density $\rho$ and depth $h_0$ bounded by a solid boundary below and an infinitely long layer of perfectly dielectric gas with permittivity $\epsilon_d$ above in a two-dimensional Cartesian $x$-$y$ coordinate system. The gravity acts in the negative $y$-direction. The interface between the fluid and the gas is free to move and is usually called a free surface. Without losing generality, we set the undisturbed free surface at $y=0$ and the bottom boundary at $y=-h_0$. Electric fields $E$ are active in the vertical direction. In the upper layer occupied by the dielectric gas, the induced magnetic field is negligible so that the electric fields admit a potential function $V(x,y,t)$, i.e. $E=\nabla V$, and satisfies $V\sim E_0 y$ as $y\rightarrow\infty$ where $E_0$ is a constant. Its irrotational nature also implies that $V$ satisfies the Laplace equation in the gas layer. Meanwhile, there is no variation in the electric potential within the fluid bulk so $V$ is constant, which is assumed to be zero without losing generality in the lower layer due to the conducting nature of the fluid. A schematic is displayed in Figure \ref{fig:scm}.  We consider a travelling wave, whose profile is described by $\zeta(x,t)$, propagating in the positive $x$-direction.  The velocity field in the bulk of fluid is  denoted by $\left(u\left(x,y,t\right),v\left(x,y,t\right)\right)$.  We denote  $l$ by a typical horizontal length scale and $a$  by a typical wave amplitude. Following \cite{GHPV}, the dimensionless variables are defined by 
 \begin{eqnarray}
 x=l x '\,,\quad  t=\frac{l t'}{c_0}\,,\quad V=E_0 l  V'\,, \quad
y^- =h_0y'\,,\quad y^+=\lambda \tilde y' \,, \quad \eta=a \eta'\,,
 \end{eqnarray} in which $c_0=\sqrt{gh_0}$ is the long-wave speed, $y^+$ and $y^-$ are the ordinates in the upper and lower layer respectively. The primes are dropped to ease the notations. We follow to introduce two parameters as follows
 \begin{equation}
     \alpha=\frac{a}{h_0}\,,\quad \beta=\frac{h_0^2}{\lambda^2}\,
 \end{equation}
 to measure amplitude and depth.  In the dimensionless variables, the bottom boundary is at $y=-1$ and the free surface is at $y=\alpha \eta(x,t)$ or $\tilde y=\alpha\sqrt{\beta}\,\eta(x,t)$.
It follows that the dimensionless governing equations are written by
	\begin{align} \label{Eq1}
		\begin{split}
			& V_{xx}+V_{\tilde y \tilde y}= 0, \;\  \mbox{for} \;\  y > \alpha\eta(x,t), \\
			& V_{x}+\alpha\sqrt{\beta}\eta_{x}V_{\tilde y}=0 , \;\  \mbox{at} \;\   y = \alpha\eta(x,t), 
   		\end{split}
	\end{align} and
   	\begin{align} \label{Eq2}
		\begin{split}
			&  u_{t}+uu_{x}+vu_{y}=-p_x  \;\  \mbox{for} \;\   -1 < y < \alpha\eta(x,t), \\
			& \beta\Big(v_{t}+uv_{x}+vu_{y}\Big)=-p_{y} \;\ \mbox{at} \;\ y =\alpha\eta(x,t),\\
			& \beta v_{x}-u_{y}=\Omega,  \;\ \mbox{for} \;\  -1 < y < \alpha\eta(x,t), \\
			& u_{x}+v_{y}=0, \;\ \mbox{for} \;\  -1 < y < \alpha\eta(x,t), \\
			&\eta_{t}+u\eta_{x}=\frac{v}{\alpha}, \;\ \mbox{at} \;\ y = \alpha\eta(x,t), \\
		\end{split}
	\end{align}
	where $\Omega=h\omega/c_0$ is the dimensionless vorticity. The Young-Laplace equation at the free surface reads
	\begin{equation}
\label{Eq3}	p-\alpha\eta -\frac{F_{E}^{2}}{2}=-\frac{F_{E}^{2}}{1+\alpha^{2}\beta(\eta_x)^{2}}\Big[\alpha^{2}\beta(\eta_x)^{2}T_{11}-2\alpha\sqrt{\beta}\eta_{x}T_{12}+T_{22}\Big]-B\alpha\beta\frac{\eta_{x}^{2}}{(1+\alpha^{2}\beta(\eta_x)^{2})^{3/2}},
	\end{equation}
	where $T$ is the Maxwell stress tensor given by 
\begin{equation}
     T_{11}=\frac{V_x^2-V_{\tilde y}^2}2=-T_{22}\,,\quad T_{12}=V_xV_{\tilde y}\,,
\end{equation} and
	\begin{equation}
	F_{E}^{2}=\frac{\epsilon_{d}E_{0}^{2}}{\rho g h}\,,\quad B =\frac{\sigma}{\rho g h^{2}},
	\end{equation}
	are called the electric Froude number and the Bond number respectively. The former parameter measures the ratio of the strength of the  electric field over gravity, and the latter measures the ratio of the capillary force over gravity. 	It is assumed that the flow is in the presence of a depth-dependent imposed current $(U(y),0)$, which dominates the velocity field. This work concerns investigating the appearance of stagnation points in the bulk of the fluid beneath a solitary wave. To this end, we first derive an asymptotic model for the velocity field and the free surface in the long-wave limits which is the so-called Korteweg-de Vries Benjamin-Ono Equation \cite{GHPV}. As the derivation is well acknowledged, we only present the main results. The readers may refer to  Hunt and Dutykh \cite{Hunt:2020} for more details.

 In the KdV scaling, we select $\alpha = \beta = \epsilon\ll 1$. For a travelling-wave solution, the variables become $\xi = x - ct$ and $\tau = \epsilon t$ in which $c$ is the linear phase speed. We seek an asymptotic solutions of (\ref{Eq1})-(\ref{Eq3}) in the form of
%	\begin{align} \label{Eq2}
%		\begin{split}
%			& V_{\xi\xi}+V_{yy}= 0, \;\  \mbox{for} \;\  y > \epsilon\eta, \\
%		& V_{x}+\epsilon^{3/2}\eta_{x}V_y=0 , \;\  \mbox{at} \;\   y = \epsilon\eta, \\
%		& \epsilon u_{\tau}-cu_{\xi}+\epsilon u u_{\xi}+\epsilon vu_{y}=-p_x = 0, \;\  \mbox{for} \;\   -1 < y < \epsilon\eta, \\
%			& \epsilon\Big(-c v_{\xi}+\epsilon v_{\tau}+\epsilon uv_{\xi}+\epsilon vu_{y}\Big)=-p_{y} \;\ \mbox{at} \;\ y =\epsilon\eta,\\
%			& \epsilon v_{\xi}-u_{y}=0,  \;\ \mbox{for} \;\  -1 < y < \epsilon\eta, \\
%		& u_{\xi}+v_{y}=0, \;\ \mbox{for} \;\  -1 < y < \epsilon\eta, \\
%			&\epsilon\eta_{\tau}-c\eta_{\xi}+\epsilon u\eta_{x}=v, \;\ \mbox{at} \;\ y = \epsilon\eta\,, \\
%			\end{split}
%	\end{align}
%	The Young-Laplace equation at the free surface reads
%	\begin{equation} \label{Eq2b}
%	 p-\eta -\frac{F_{E}^{2}}{2\epsilon}=-\frac{1}{\epsilon}\frac{F_{E}^{2}}{1+\epsilon^{3}(\eta_\xi)^{2}}\Big[\epsilon^{3}(%\eta_\xi)^{2}T_{11}-2\epsilon^{3/2}\eta_{\xi}T_{12}+T_{22}\Big]-B\epsilon\frac{\eta_{\xi}^{2}}{(1+\epsilon^{3}(\eta_\xi)^{2})^{3/2}}.
%	\end{equation}

%Now, we introduce  the new scaling 	
%	\begin{align}\label{s1} 
%		\begin{split}
%			u\rightarrow -\Omega y+ \epsilon u, \;\ v \rightarrow \epsilon v, \;\ p \rightarrow\epsilon p, 
%		\end{split}
%	\end{align} 
%	and seek for asymptotic solutions of (\ref{Eq1})-(\ref{Eq3}), we define the power series expansion
\begin{align} \label{expansions}
		\begin{split}
			& u(\xi,y,\tau)= -{\Omega y} + \epsilon u_{0}(\xi,y,\tau) + \epsilon^2 u_{1}(\xi,y,\tau)+{o}(\epsilon^2) \\
			& v(\xi,y,\tau)= \epsilon v_{0}(\xi,y,\tau) + \epsilon^2 u_{1}(\xi,y,\tau)+{o}(\epsilon^2) \\
			& p(\xi,y,\tau)=\epsilon p_{0}(\xi,y,\tau) + \epsilon^2 p_{1}(\xi,y,\tau)+{o}(\epsilon^2) \\
			& V(\xi,y,\tau)= -y + \epsilon^{3/2}V_{1}(\xi,y,\tau) +{o}(\epsilon^{3/2})	\\
			& \eta(\xi,\tau)= \eta_{0}(\xi,\tau) + \epsilon \eta_{1}(\xi,\tau)+{o}(\epsilon)
			\end{split}
	\end{align}
 Substituting  (\ref{expansions}) in equations  (\ref{Eq1})-(\ref{Eq3}), it is discovered at the leading order that
	\begin{equation}\label{Burns1}
		c^2 -\Omega c =1.
	\end{equation}
Here, we choose the solution with a positive sign, i.e. 	$c =\frac{\Omega}{2}+\frac{\sqrt{\Omega^{2}+4}}{2}$.  We note that \eqref{Burns1} is in fact the linear dispersion relation in the long-wave limit, i.e. when $k\rightarrow 0$, and therefore the surface tension and the electric fields do not contribute. At the quadratic order, a KdV-Benjamin-Ono equation that incorporates the surface tension, vorticity effects and electric forces is obtained to be
\begin{equation}\label{KdV}
\eta_{0\tau}+\mu\eta_{0}\mathbf{\eta}_{0\xi}+\nu{\eta}_{0\xi\xi\xi}+\gamma\mathcal{H}[\eta_{0\xi\xi}]=0
\end{equation}
where the coefficients are given by
\begin{equation}\label{coefficients}
\mu=\frac{\Omega^{2}+3}{2c-\Omega}, \;\ \nu=\frac{1}{2c-\Omega}\left(\frac{c^2}{3}-B\right), \;\ \gamma=-\frac{F_{E}^2}{2c-\Omega}.
\end{equation}
and $\mathcal{H}$ is the Hilbert operator which is defined as
\begin{equation}\label{Hilbert}
\widehat{\mathcal{H}[f(\xi)]}=-i\sign(k)\hat{f}(k),
\end{equation}
where $\widehat{[\cdot]}$ represents the Fourier transform.

In the absence of electric forces ($\gamma=0$), equation \eqref{KdV} reduces to a standard KdV equation which admits solitary wave solutions described by the formula \cite{Whitham:1974}
\begin{equation}
\eta_{0}(\xi,\tau)=A\sech^{2}\left(\sqrt{\frac{\mu A}{12\nu}}\left(\xi-\frac{\mu A}{3}\tau\right)\right).
\end{equation}
It is noted that the KdV equation collapses  when the nonlinearity disappears at $\nu=0$, or at
\begin{equation}
    B = B_c\equiv\frac{c^2}{3}\,.
\end{equation}
Under such circumstances, a different scaling is required to derive a fifth-order KdV equation. As it is irrelevant to the main aim of this work, it will not be further discussed. When $\nu$ is non-zero (and $\gamma=0$), equation (\ref{KdV}) admits elevation solitary wave solutions  when $0\le B<B_c$ and depression solitary wave solutions when $B>B_c$.

When electric forces are present, solitary waves of equation (\ref{KdV}) do not have a closed form. We consider a solitary wave solution  of (\ref{KdV}) denoted by  $\Theta=\Theta(\xi-C\tau)$  propagating with speed $C$. It immediately follows that $\eta_{0}(\xi,\tau)=\Theta(\xi-C\tau)$. As we are interested in investigating  particle trajectories for the Euler equations using the KdV-Benjamin-Ono model as an approximation, we have to express the free surface and the horizontal velocity at the bottom of the channel using the Euler coordinates. The solitary wave solution  and the approximation of the velocity field  in the Euler coordinates are written respectively by
\begin{equation}\label{solitaryeuler}
\eta_{0}(x,t)=\Theta(x-(c+\epsilon C)t),
\end{equation}
and 
\begin{equation}\label{velocityeuler}
u_{0}(x,y,t)=c\Theta(x-(c+\epsilon C)t)) \mbox{ and } v_{0}(x,y,t)=-c\Theta_{x}(x-(c+\epsilon C)t))(y+1).
\end{equation}
In the next section, we present the numerical methods to compute depression solitary waves of (\ref{KdV}) to investigate particle trajectories in the bulk of the fluid.

\section{Numerical methods}\label{ns}
%\begin{figure}[!h]
%	\centering	
%	\includegraphics[scale =1]{Fig1e.pdf}
%	\caption{Wavepacket solutions of equation (\ref{KdV}). Parameters: $B=10$ and $\Omega=5$.}
%	\label{Fig1b}
%\end{figure}

Solitary waves $\Theta$ with speed $C$, amplitude $A$ and crest located at $x=0$ for the KdV-Benjamin-Ono equation (\ref{KdV}) are computed through Newton's method by solving the equations
\begin{align}\label{Newton}
\begin{split}
-C\Theta_{\xi}+\alpha\Theta\Theta_{\xi}+\beta\Theta_{\xi\xi\xi}+\gamma\mathcal{H}[\Theta_{\xi\xi}]=0, \\
\Theta(0)-A=0,
\end{split}
\end{align}
in a periodic computational domain $[-L,L)$  with a uniform grid with  even points $N$. The spatial points are discretised as
\begin{equation}\label{grid}
\xi_j = -L + (j-1)\Delta\xi\,,\, \mbox{for $j=1,2,\dots$, $N$,\, where $\Delta\xi = 2L/N$,}
\end{equation}
and the frequencies as
\begin{equation}\label{freq}
(k_1,k_2,\cdots,k_N) = \frac{\pi}{L}(0,1,\cdots,N/2-1,0,-N/2+1,\cdots,-1).%k_j =(-\frac{N}{2}+  j)\frac{\pi}{L},  \mbox{ } j=1,2,\dots,N
\end{equation}

On the grid points defined in equation (\ref{grid}), we denote by $\Theta_j=\Theta(\xi_j)$, $\Theta_{\xi,j}=\Theta_\xi(\xi_j)$, $\Theta_{\xi\xi,j}=\Theta_{\xi\xi}(\xi_j)$ and  $\Theta_{\xi\xi\xi,j}=\Theta_{\xi\xi\xi}(\xi_j)$. The discretised version of equations ({\ref{Newton}})  gives rise to a system of ($N+1$) equations  with ($N+1$) unknowns
\begin{align}\label{EE3}
\begin{split}
G_{j}(\Theta_1,\Theta_2,...,\Theta_{N},C):=-C\Theta_{\xi, j}+\alpha\Theta_{j}\Theta_{\xi, j}+\beta\Theta_{\xi\xi\xi, j}+\gamma\mathcal{H}[\Theta_{\xi\xi, j}]=0,  \mbox{ for $j=1,2,\dots$, N.} \\
G_{N+1}(\Theta_1,\Theta_2,...,\Theta_{N},C):=\Theta_{N/2+1}-A=0.
\end{split}
\end{align}
The discretisation chosen allows us to compute all spatial derivatives and the nonlocal  operator $\mathcal{H}$ in equations (\ref{Newton}) with spectral accuracy in Fourier space through the FFT \cite{Trefethen:2001}. The system's Jacobian for the Newton iteration is found by finite variations in the unknowns and the stopping criterion considered is
\begin{align}
\begin{split}
\frac{\sum_{j=1}^{N+1}|G_{j}(\Theta_1,\Theta_2,...,\Theta_{N},C)|}{N+1}< \delta,
\end{split}
\end{align}
where $\delta$ is the tolerance value set to be $10^{-10}$. For a fixed value of $A$, $\Omega$ and $F_{E}$, we choose  the solitary wave solution of  equation (\ref{KdV}) in the absence of electric forces 
\begin{equation}\label{Guess}
\Theta_{0}(\xi) = A\sech^{2}(k\xi)\,,\quad C_0 = -\frac{\alpha A}{3}.
\end{equation}
  as the initial guess. The solution is then computed by a continuation method in the parameter $F_{E}$ by using the prior converged solution of the Newton method as the initial guess.
  \begin{figure}[!h]
	\centering	
	\includegraphics[scale =1]{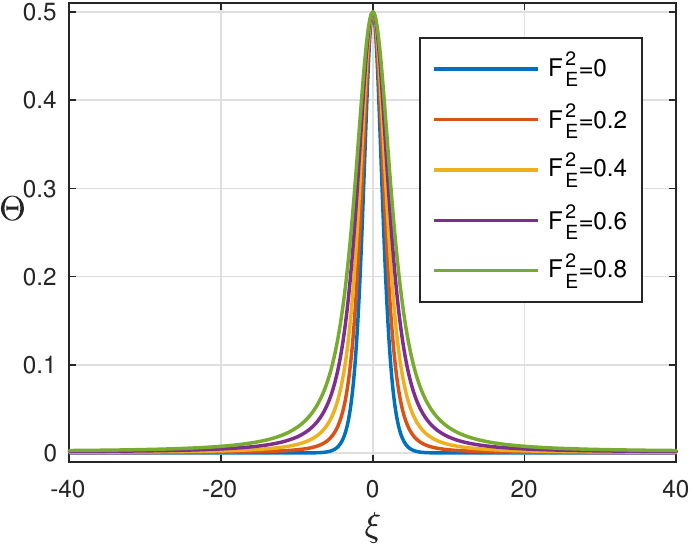}
	\includegraphics[scale =1]{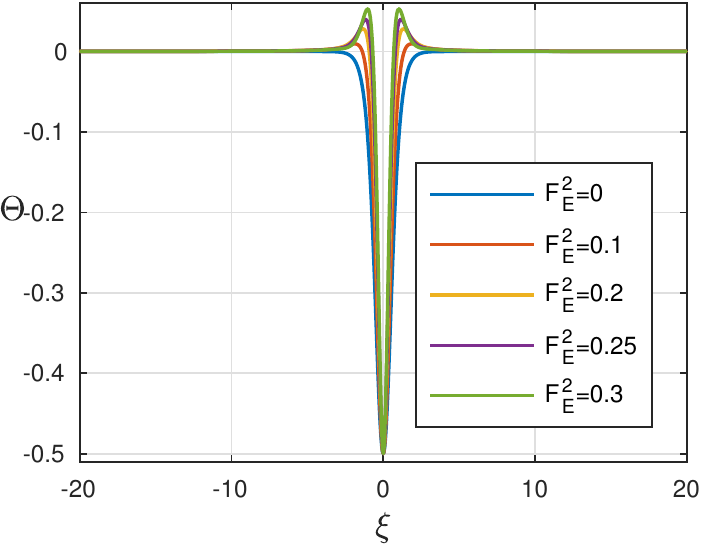}
	\includegraphics[scale =1]{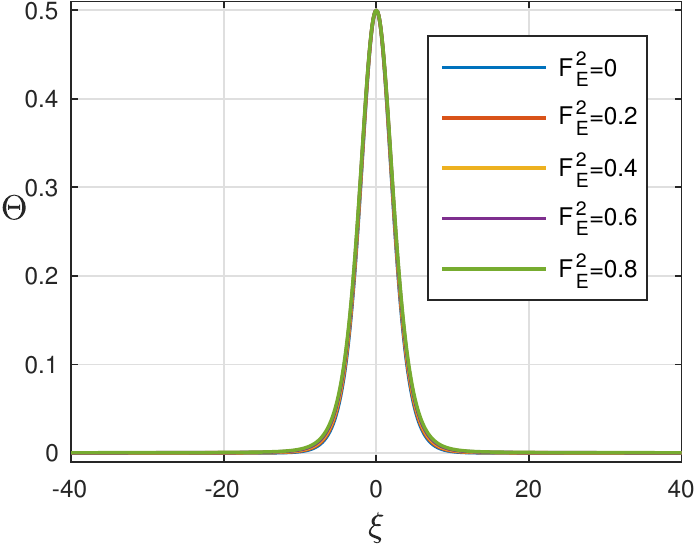}
	\includegraphics[scale =1]{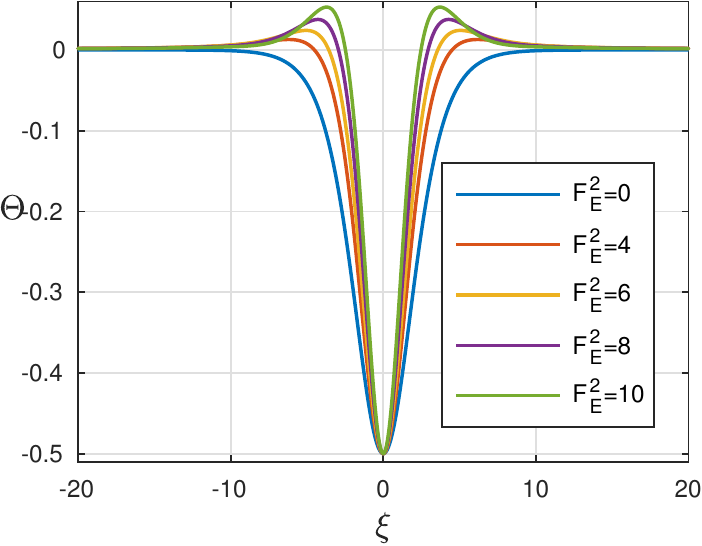}
	\caption{Top: Solitary wave solutions of equation (\ref{KdV}) in the absence of vorticity ($\Omega=0$) and different values of $F_E$. Parameters: $B=0$ (left) and $B=0.4$ (right). Bottom: Solitary wave solutions of equation (\ref{KdV}) with $\Omega=5$ and different values of $F_E$. Parameters: $B=0$ (left) and $B=17$ (right).}
	\label{Fig1}
\end{figure}

Typical numerical solitary waves are displayed in Figure \ref{Fig1}. We recall that elevation solitary waves occur when $B<B_c$ and depression ones when $B>B_c$. 
It is noted that the elevation and depression solitary waves have more ripples appearing on the side of the main pulse when the third-order dispersive term is weak and the electric term in the Hilbert transform is strong, i.e. $\nu$ small and $\gamma$ big in \eqref{KdV}. These solutions with decaying oscillatory tails have been previously reported by \cite{Albert:2014} and \cite{DDM}. For the purpose of this work, we only focus on the waves shown in Figure \ref{Fig1}.

%%%%%%%%%%%%%%%%%
\section{Particle trajectories}\label{pt}
Particle trajectories beneath the solitary wave (\ref{solitaryeuler}) can be computed approximately by solving the dynamical system 
\begin{align} \label{DSf}
\begin{split}
& \frac{dx}{dt} =-\Omega y + \epsilon u(x,y,t)\approx -\Omega y  + \epsilon c\Theta(x-(c+\epsilon C)t)) , \\
& \frac{dy}{dt} =\epsilon v(x,y,t)\approx-\epsilon c\Theta_{x}(x-(c+\epsilon C)t))(y+1). \\
\end{split}
\end{align}
%%% Deveria ser eps dx/dt = ... e eps dy/dt = ... por causa do tau na mudanca de variavel, mas fica confuso escrever aqui!!! Por isso que no 
%%% codigo a funcao de corrente aparece dividida por eps.

In order to compute stagnation points, it is convenient to solve equations (\ref{DSf}) in the frame that moves with the wave speed, for this purpose we consider the new variables $X= x-(c+\epsilon C)t$ and $Y=y$. In this new reference frame, the streamlines are solutions of the autonomous dynamical system
\begin{align} \label{DS}
\begin{split}
& \frac{dX}{dt} = - \Omega Y + \epsilon c\Theta(X) -(c+\epsilon C) , \\
& \frac{dY}{dt} =-\epsilon c\Theta_{X}(X)(Y+1), \\
\end{split}
\end{align}
which can be seen as the level curves of the Hamiltonian $\mathbf{\Psi}(X,Y)$  given by
\begin{equation} \label{stream function}
\mathbf{\Psi}(X,Y) = \epsilon c\Theta(X)(Y+1)-\frac{\Omega}{2}Y^{2}-(c+\epsilon C)Y.
\end{equation}
Notice that once the solitary wave $\Theta$ is computed numerically through the method proposed in the previous section, the level curves  can be easily computed using the function {\it contour} that is implemented in {MATLAB}.

In the absence of surface tension and electric fields, Guan \cite{Guan:2020} investigated particle trajectories beneath solitary waves in the presence of a linear sheared current through the Korteweg-de Vries equation. He showed that the orbits obtained from the asymptotic approximation agree well with the ones computed through the full Euler equations when the solitary waves have  small amplitudes. Based on his results, in all simulations presented in this article, we fix $\epsilon =0.1$.
\section{Results and discussion}\label{res}
\subsection{Elevation solitary waves}

In the absence of an electric field, the increase of the vorticity could cause the appearance of stagnation points (see \cite{Ribeiro-Jr:2017}).  It first appears at the bottom and below the crest. As the vorticity increases further, other stagnation points appear in the bulk of the fluid creating a recirculation zone \cite{Ribeiro-Jr:2017}. Therefore, in order to discuss the influence of the electric field in the flow structure beneath solitary waves for $0 \le B < B_c$, we first find the smallest value of the vorticity such that a stagnation point appears at the bottom and below the solitary wave crest in the absence of the electric field then follow to study  the case where the electric fields are switched on. 

The value of the vorticity for which we have a single  stagnation point located at  the bottom and below the solitary wave crest is obtained by solving for $\Omega$ equation (\ref{DS}) evaluated at $X=0$ and $Y=-1$ which yields the equation
\begin{equation}\label{stag1}
0=\Omega +\epsilon cA -(c+\epsilon C).
\end{equation}
The solution to equation  (\ref{stag1}) for $F_{E}=0$ and $B=0$ is $\Omega^{*}\approx 5.2962$ and this value does not vary considerably with $B$ because as pointed out by Flamarion \cite{TEMA:2023} surface tension does not create stagnation points. Moreover, it barely changes the position of the stagnation point below the crest (when it does exist). 
\begin{figure}[!h]
	\centering	
	\includegraphics[scale =1]{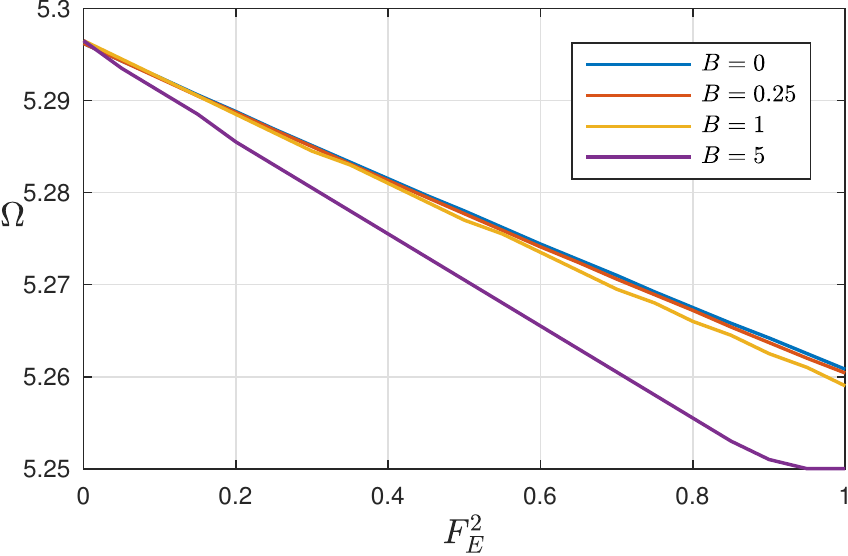}
	\caption{The graph represents the vorticity as a function of the parameter $F_E$ in which the first stagnation point gives rise at the bottom of the channel.}
	\label{Fig2}
\end{figure}

\begin{figure}
	\centering	
	\includegraphics[scale =1]{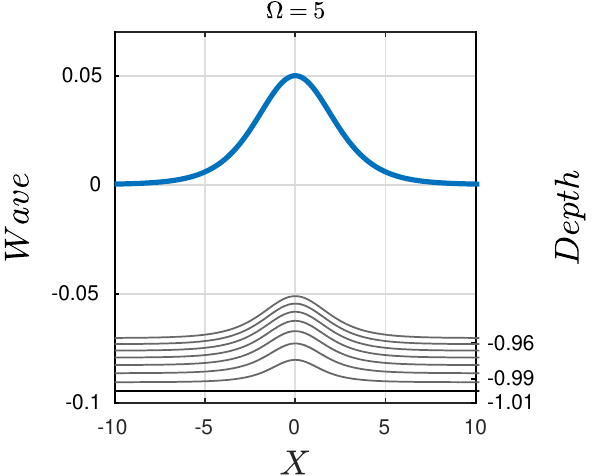}
	\includegraphics[scale =1]{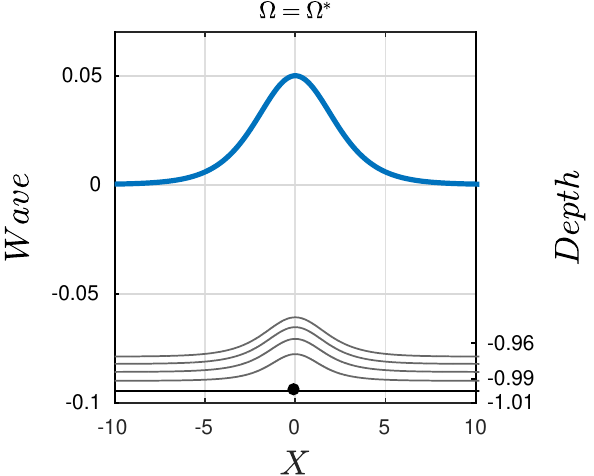}
	\includegraphics[scale =1]{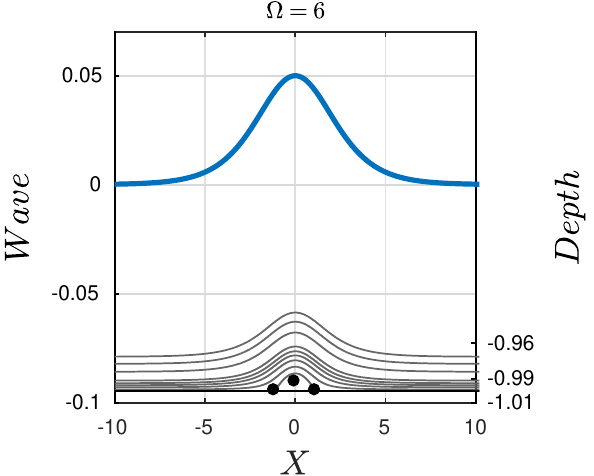}
	\includegraphics[scale =1]{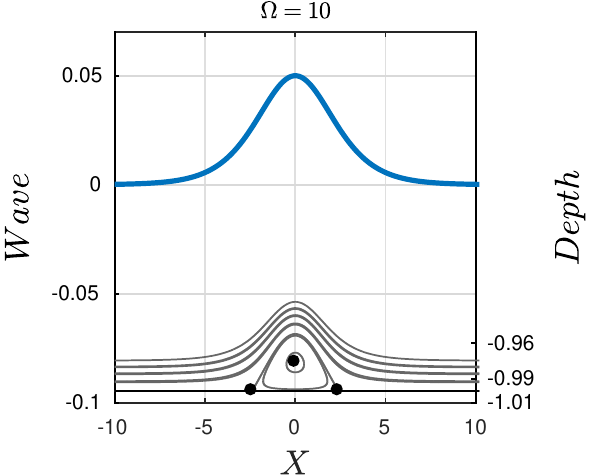}
	\caption{Phase portrait for different values of the vorticity with a solitary wave with amplitude $A=0.5$, $F_{E}^2=0.5$ and $B=0$. The critical value of the vorticity in which the first stagnation point appears at the bottom is $\Omega^{*}\approx 5.2780$.}
	\label{Fig3}
\end{figure}

Figure \ref{Fig2}  displays the solution of equation (\ref{stag1}) for different values of the Bond number.   These curves correspond to flows   with a single stagnation point on the bottom and beneath the crest.  Firstly, it is noticed that the solution does not vary much for different values of the Bond number and small values of the parameter $F_E$. Besides, we observe that the appearance of the stagnation point on the bottom can occur at a tinnier vorticity with the increase of intensity of the electric field. Secondly,  we can  regard these curves as   bifurcation points that separate the  parameter space in two regions according to the number of stagnation points beneath the solitary wave. For those $(F_{E}^2,\Omega)$  below these curves,  there is no stagnation point in the fluid domain. On the other side, for those  $(F_{E}^2,\Omega)$   above these curves,  there exist three stagnation points, namely, two saddles at the bottom of the channel and a centre in the bulk of the fluid aligned with the crest of the solitary wave. And there is only one stagnation point at the bottom for those  $(F_{E}^2,\Omega)$ right on the curves. A typical example of this bifurcation is depicted  in Figure \ref{Fig3}.

\begin{figure}[!h]
	\centering	
	\includegraphics[scale =1]{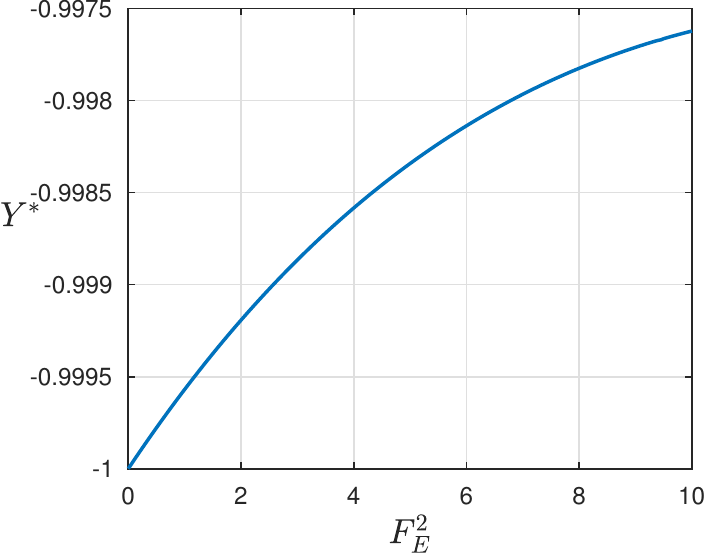}
	\includegraphics[scale =1]{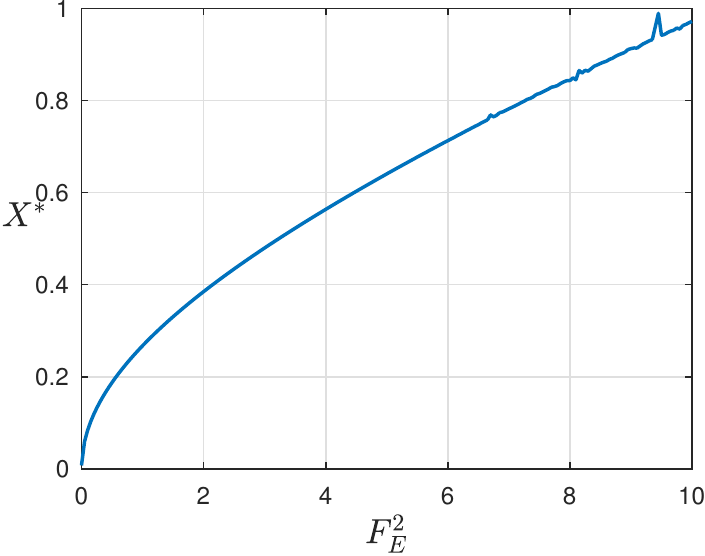}
	\caption{The effect of the electric field in the position of the centre bellow the crest of the solitary wave for $B=0$ and $\Omega=\Omega^{*}$.}
	\label{Fig4}
\end{figure}

We follow to analyse how the strength of the electric field affects the location of the stagnation points. To this end, we fix the vorticity and the surface tension and let $F_E$ vary.  The left panel of Figure \ref{Fig4} shows the vertical position ($Y^*$) of the stagnation point located below the wave crest and the right panel of the same figure presents the horizontal coordinate ($X^*$) of the saddle point as a function of the parameter   $F_E$  for $\Omega=\Omega^{*}$ and $B=0$. Of note, the intensity of $F_E$ barely impacts the position of the centre point, however, it does affect the position of the saddle points. 

Flamarion {\it et al.} \cite{FGRD:2022} showed that the appearance of stagnation points beneath periodic travelling waves can occur at small vorticity with the help of electric fields. Besides, it was shown that the position of all  the stagnation points changed significantly with variations in the electric field. The features differ from the discussion presented above for elevation solitary waves where the electric field does not act as a mechanism to help the generation of stagnation points. 

\subsection{Depression solitary waves}
It is known that in the absence of an electric field typical depression solitary wave solutions of equation (\ref{KdV}) are $\sech^{2}-$like. For such waves, it is well established  that stagnation points never give rise to the bulk of the fluid. As shown by Flamarion \cite{Wave Motion:2023} stagnation points beneath depression solitary waves can occur only in the presence of decaying or oscillatory tails, which cannot be captured by a third-order KdV equation. As can be seen from Figure \ref{Fig1}, under the electrical effect, equation (\ref{KdV}) admits depression solitary wave solutions with two elevation dimples on the side of the wave trough. Consequently, an immediate interesting question is whether stagnation points can take place in the bulk of the fluid beneath such depression solitary waves, which will be examined in the rest of the paper. %

Other authors have studied the appearance of stagnation points beneath depression solitary waves \cite{Wave Motion:2023, TEMA:2023,  Wang:2020}. However, these works considered gravity-capillary waves in the absence of electric fields. Moreover, it has been shown that the location of the stagnation points does not change much for choices of $B$. Having said this, we focus on investigating the effect of the electric field as a mechanism to create stagnation points. To address this issue we fix the vorticity, the Bond number and vary the intensity of the electric field. 

\begin{figure}
	\centering	
	\includegraphics[scale =1]{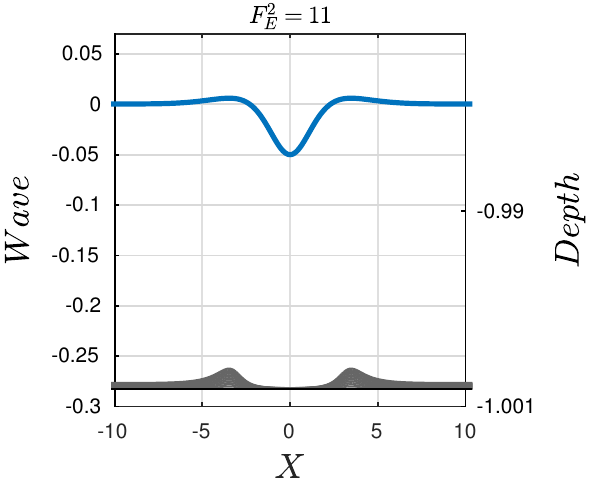}
	\includegraphics[scale =1]{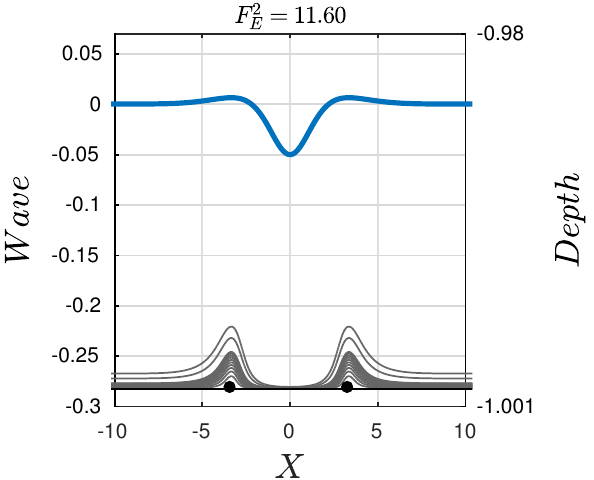}
	\includegraphics[scale =1]{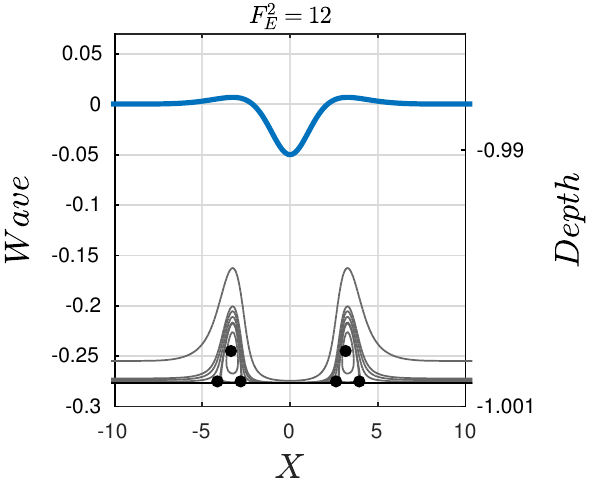}
	\includegraphics[scale =1]{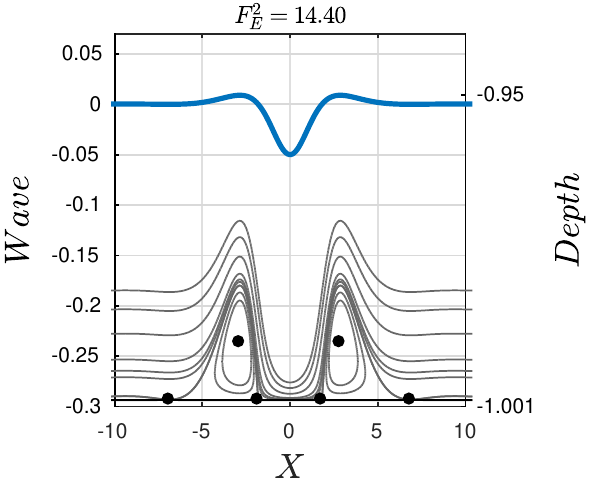}
	\includegraphics[scale =1]{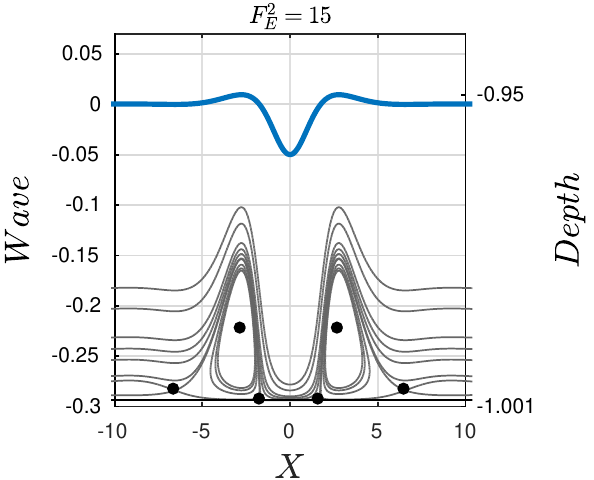}
	\includegraphics[scale =1]{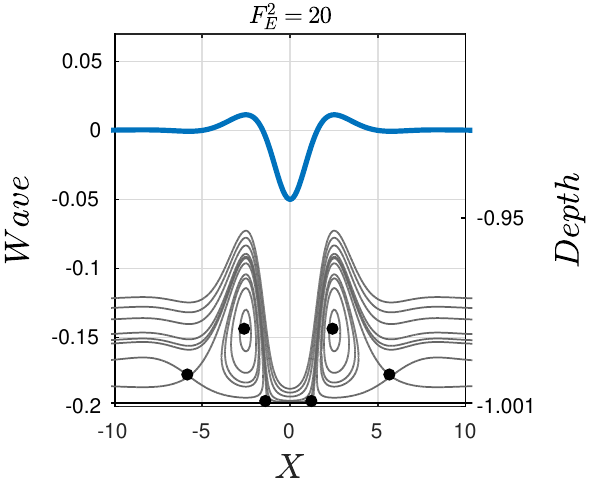}
	\caption{Phase portrait for different values of the vorticity with a solitary wave with amplitude $A=0.5$, $\Omega=5$ and $B=17$. The critical value of the vorticity in which the first stagnation point appears at the bottom is $\Omega^{*}\approx 5.2780$.}
	\label{Fig5}
\end{figure}

Figure \ref{Fig5} depicts a series of simulations from which we can see that in the presence of a strong electric field stagnation points can appear in the fluid domain.   The location of the stagnation points is determined in two ways-- (i) by finding the equilibrium points of the dynamical system (\ref{DS}), i.e., we find the zeros of the velocity field or (ii) by the contour function of MATLAB.  The flow structure beneath the depression solitary wave can have (i) zero, (ii) two centres (at the bottom), (iii) two centres (in the bulk of the fluid) and four saddles (at the bottom) or (iv) two centres (in the bulk of the fluid) and four saddles (two at the bottom and two in the bulk of the fluid) as stagnation points depending on the intensity of the electric field. This features the bifurcation of flow according to the $F_{E}^2$ parameter. Similar descriptions of the arrangement of the stagnation point in the context of gravity-capillary waves were reported in the work of Flamarion \cite{Wave Motion:2023}.

It is well acknowledged that the full Euler equations are the most realistic model to reproduce EHD scenarios in inviscid fluids. However, reduced models can reproduce qualitatively the same features of the flow with comparatively little effort.  For instance, our results show that the weakly nonlinear weakly dispersive regime can capture rich flow structures, such as recirculation zones and stagnation points.

	\section{Conclusion}\label{conclusion}
In the presented study  the flow structure beneath  EHD flows with constant vorticity was investigated numerically in the Korteweg Benjamin-Ono equation framework. Solitary waves were computed numerically through the standard Newton's method combined with Fourier spectral methods. This approach allowed us to approximate the velocity field beneath the free surface. As a consequence, the location of stagnation points and details of the recirculation zones formed by them were determined. For elevation solitary waves, we showed that the location of the centre points does not change significantly by variations on the electric field. It is remarkable that for depression solitary waves the electric field acts as a mechanism for the creation of stagnation points. In the absence of an electric field even when the vorticity is strong there is no stagnation point in the bulk of the fluid. The results presented in this work are expected to agree well with the full nonlinear model. An attempt to compare the results predicted by both models is a natural path to be pursued in future.
	
%	In this paper, we  performed a quantitative study on the effects	of normal electric fields in the flow structure beneath  periodic capillary-gravity travelling waves with constant vorticity. By fixing the vorticity, it was discovered that the flow can have zero, one, two or three stagnation points depending on the value of $E_b$. The appearance of a single stagnation point was found to occur at $E_b=E_{b1}$.	When $E_b>E_{b1}$, there exist multiple stagnation points which form a  recirculation zone. At $E_b=E_{b2}$, two saddle points  from consecutive periods collide and merge into one which shifts away from the bottom boundary with a further increase of $E_b$. More interestingly,  the impact of the electric effect on the pressure beneath the surface was shown to be significant. 	Besides, we demonstrated that the locations of  local maxima for the pressure on the lower boundary are intrinsically associated to the projections of the saddle points onto the bottom. To the best of our knowledge, the study carried out in this paper is the first one in the literature devoted to the investigation of stagnation points and pressure anomalies in the context of electrohydrodynamics  flows. It opens the door to other fronts of study in the such field which eventually can lead to investigating more realistic flows. 

%	
	\section*{Acknowledgments}
	 M. V. F  and R.R.-Jr are  grateful to IMPA for hosting them as visitors during the 2023	 Post-Doctoral Summer Program.

	\section*{Data Availability Statement}
	Data sharing is not applicable to this article as the parameters used in the numerical experiments are informed in this paper.


\begin{thebibliography}{}
\bibitem{CCY}
	X. Chen, J. Cheng and X. Yin. Advances and applications of electrohydrodynamics, {\em Chin. Sci. Bull}. \textbf{48} (2003) 1055–1063.

 \bibitem{D}
 D. T. Papageorgiou, Film flows in the presence of electric fields, {\em Ann. Rev. Fluid Mech.} \textbf{51} (2019) 155–187.

\bibitem{DGVK}
 A. Doak, T. Gao, J.-M. Vanden-Broeck \& J. J. S. Kandola, Capillary-gravity waves on the interface of two dielectric fluid layers under normal electric fields, {\em Q. J. Mech. Appl. Math.} \textbf{73} (2020) 231–250.

\bibitem{Wang}
 Z. Wang, Modelling nonlinear electrohydrodynamic surface waves over three-dimensional conducting fluids, {\em Proc. R. Soc.} A \textbf{473} (2017) 20160817.
 
  \bibitem{FGRD:2022}
 M. V. Flamarion,  T. Gao,  R. Ribeiro-Jr \&  A. Doak.
 Flow structure beneath periodic waves with constant vorticity under normal electric fields.
 \emph{Phys. Fluids} \textit{34}, 127119 (2022). 

  \bibitem{GHPV} H. Gleeson, P. Hammerton, D. Papageorgiou, J.-M. Vanden-Broeck, A new application of the Korteweg-de Vries Benjamin-Ono equation in interfacial electrohydrodynamics, \emph{Phys. Fluids} \textbf{19} (2007) 031703.
 
 \bibitem{BK}
 H. Borluk, H. Kalisch, Particle dynamics in the KdV approximation, Wave Motion 49 (2012) 691-709.
 
 \bibitem{G}
 L. Gagnon, Qualitative description of the particle trajectories for n-solitons solution of the korteweg-de Vries equation, Discrete Contin. Dyn.
Syst. 37 (2017) 1489-1507.

\bibitem{Khorsand}
Z. Khorsand, Particle trajectories in the Serre equations, Appl. Math. Comput. 230 (2014) 35-42.


 	 \bibitem{Guan:2020}
	{X. Guan,} 
	{Particle trajectories under interactions between solitary waves and a linear
shear current.}
	{Theor. App. Mech. Lett.} 10 (2020) 125-131.
	
	\bibitem{AK}
 A. Alfatih, H. Kalisch, Reconstruction of the pressure in long-wave models with constant vorticity, Eur. J. Mech. B Fluids 37 (2013) 187-194.
 


\bibitem{CK}
C. Cutis, J. Carter, H. Kalisch, Particle paths in nonlinear Schr\"odinger models in the presence of linear shear currents, J. Fluid Mech. 855 (2018)
%322-350.

\bibitem{CK2}
J. Carter, C. Curtis, H. Kalisch, Particle trajectories in nonlinear Schr\"dinger models, Water Waves 2 (2020) 31-57.

	\bibitem{Wave Motion:2023}
{M.V. Flamarion,} 
	{Complex flow structures beneath rotational depression solitary waves.}
	{Wave Motion.} 117, (2023) 103108.
	
		\bibitem{TEMA:2023}
{M.V. Flamarion,} 
	{Stagnation points beneath rotational solitary waves in gravity-capillary flows.}
	{Trends in Computational and Applied Mathematics.} (in press) (2023).
	
	\bibitem{FR2023}
{M.V. Flamarion,  R. Ribeiro-Jr,} 
Solitary Waves on Flows with an Exponentially Sheared Current and Stagnation Points. \emph{Quart. J. Mech. Appl. Math.}, (2023). 

 

 
 \bibitem{Hunt:2020}
 M. J. Hunt \& D. Dutykh,  Free Surface Flows in Electrohydrodynamics with a Constant Vorticity Distribution. \emph{Water Waves} \textbf{3}, 297–317 (2021). 
 
 
 \bibitem{Whitham:1974}
{G.B. Whitham, } 
\newblock{Linear and Nonlinear Waves.}
\newblock{Wiley.} 1974.

 	
 \bibitem{Trefethen:2001}{Trefethen LN} 
 {Spectral Methods in MATLAB.}
 { Philadelphia: SIAM;} 2001.

\bibitem{Albert:2014}
{J. P. Albert, J. L. Bona \& J. M. Restrepo, } 
\newblock{Solitary-wave solutions of the Benjamin equation.}
\newblock{\emph{SIAM J. Appl. Math.}} \textbf{59}(6), 2139-2161 (2014).





\bibitem{DDM}
{V. Dougalis, A. Duran, \& D. Mitsotakis,}
\newblock{Numerical solution of the Benjamin equation.}
\newblock{\emph{Wave Motion}.} \textbf{52},  194-215 (2015).

 	 \bibitem{Ribeiro-Jr:2017}
	{R. Ribeiro-Jr, P.A. Milewski, A. Nachbin,} 
	{Flow structure beneath
rotational water waves with stagnation points.}
	{J. Fluid. Mech.} 812 (2017) 792-814.
	
			\bibitem{Wang:2020}
	{Z. Wang, X. Guan \& J-M. Vanden-Broeck,} 
	{Progressive flexural-gravity waves with constant vorticity.}
	{\em J. Fluid. Mech.} 995 (2020) A12.




	
	

%\bibitem{Stokes:1847} {G.G. Stokes,} 
  %    {On the theory of oscillatory waves.}
%	{ Trans. Cambridge. Phil. Soc.} 8 (1847) 441-455.
	
%	\bibitem{Ursell:1953} {F. Ursell,} 
    %  {Mass transport in gravity waves.}
	%{ Proc. Cambridge. Phil. Soc.} 40 (1953) 145-150.
	
%		\bibitem{Constantin:2008} {A. Constantin, G. Villari,} 
 %     {Particle trajectories in linear water waves.}
%	{ J. Math. Fluid. Mech.} 10 (2008) 1336-1344.
	
		
	%	\bibitem{Constantin:2006} {A. Constantin, G. Villari,} 
  %    {The trajectories of particles in Stokes waves.}
%	{Invent. Math.} 166 (2006) 523-535.

   %		\bibitem{Constantin:2010} {A. Constantin, W. Strauss,} 
   %   {Pressure beneath a Stokes wave.}
	%{ Comm. Pure. Appl. Math.} 63 (2010) 533-557.
	
%		\bibitem{Nachbin:2014} {A. Nachbin, R. Ribeiro-Jr,} 
  %    {A boundary integral method formulation for particle trajectories in Stokes Waves.}
%	{ DCDS-A.} 34(8) (2014) 3135-3153.

%	\bibitem{Kalisch:2012} {H. Borluk, H. Kalisch,} 
    %  {Particle dynamics in the KdV approximation.}
%	{ Wave Motion.} 49 (2012) 691-709.
	
%	\bibitem{Kalisch:2013} {A. Alfatih, H. Kalisch,} 
   %   {Reconstruction of the pressure in long-wave models with constant vorticity.}
%	{ Eur. J. Mech. B-Fluid.} 37 (2013) 187-194.
	
%	\bibitem{Gagnon:2017} {L. Gagnon,} 
  %    {Qualitative description of the particle trajectories for n-solitons solution of the korteweg-de vries equation.}
%	{ Discrete. Contin. Dyn. Syst.} 37 (2017) 1489-1507.
	


	% \bibitem{Khorsand:2014}
%	{Z. Khorsand,} 
%	{Particle trajectories in the Serre equations.}
	%{Appl. Math. Comput.} 230 (2014) 35-42.
	
	
%	 \bibitem{Kalisch:2018}
%	{C. Cutis, J. Carter, H. Kalisch,} 
%	{Particle paths in nonlinear schr{\" o}dinger models in the presence of linear shear currents.}
%	{J. Fluid. Mech.} 855 (2018) 322-350.

 %\bibitem{Kalisch:2020}
%	{J. Carter, C. Curtis, H. Kalisch,} 
%	{Particle trajectories in nonlinear Schr{\" o}dinger models.}
%	{Water Waves.} 2 (2020) 31-57.

% \bibitem{Ehrnstrom:2008}
%	{M. Ehrnstr{\" o}m,  G. Villari,} 
%	{Linear water waves with vorticity: rotational features and particle
%paths.}
%	{J. Differ. Equ.} 244 (2008) 888-1909.
%	
%	 \bibitem{Wahlen:2009}
%	{E. Wahl{\' e}n,} 
%	{Steady water waves with a critical layer.}
%	{J. Differ. Equ.} 246 (2009) 2468-2483.

% \bibitem{Teles:1988}
%	{A.F. Teles Da Silva, D.H. Peregrine,} 
%	{Steep, steady surface waves
%on water of finite depth with constant vorticity.}
%	{J. Fluid Mech.} 195 (1988) 281-302.
%	
%
%
%
%		
%	\bibitem{Flamarion:2020}
%	{M.V. Flamarion, A. Nachbin, R. Ribeiro-Jr,} 
%	{Time-dependent Kelvin cat-eye structure due to
%current-topography interaction.}
%	{J. Fluid. Mech.} 889 (2020) A11.
%
%
%\bibitem{Selecciones:2021}
%{M.V. Flamarion,} 
%	{Rotational gravity-capillary waves generated by a moving disturbance.}
%	{Selecciones Matem{\' a}ticas.} 8(2) (2021) 228-234.
%	
%\bibitem{Fracius:2016}
%{M. Fracius, H.C. Hsu, C. Kharif, P. Montalvo,} 
%	{Gravity-capillary waves in finite depth on flows of constant vorticity.}
%	{Proc. R. Soc. Lond A.} 472(20160363) (2016).
		



%	\bibitem{Milewski:1998}
%	{P.A. Milewski, J-M. Vanden-Broeck,} 
%	{Time dependent gravity-capillary flows past an obstacle.}
%	{Wave Motion.} 29 (1998) 63-79.
	
%{	
%	\bibitem{Hunter:1988}
%	{J.K. Hunter, J Scheurle,} 
%	{Existence of perturbed solitary wave solutions to a model equation for water
%waves.}
%	{Physica D.} 32 (1988) 253-268.}
	





%	\bibitem{Zufiria:1987}
%	{J.A. Zufiria,} 
%	{Symmetry breaking in periodic and solitary gravity-capillary waves on water of finite depth.}
%	{J. Fluid. Mech.} 184 (1987) 183-206.

%\bibitem{Toland:1993}
%	{A.R. Champneys, J.F. Toland,} 
%	{Bifurcation and coalescence e of a plethora of large amplitude homoclinic orbits for Hamiltonian systems.}
%	{Nonlinearity.} 6 (1993) 665-721.


	
%	{
%		\bibitem{Malomed:1996}
%	{B. Malomed, J-M. Vanden-Broeck,} 
%	{Solitary wave interaction for the fifth-order KdV equation.}
%	{\em Contem Math.}  200 (1996) 133-143.}





 
 


	
	
	
	
\end{thebibliography}
	\end{document}